\documentclass[runningheads]{llncs}
\usepackage[T1]{fontenc}
\usepackage{graphicx}
\usepackage{hyperref}
\usepackage{color}

\usepackage{amsmath}
\usepackage{amssymb}
\usepackage{cleveref}
\usepackage{booktabs}
\usepackage{rotating}
\usepackage{float}

\definecolor{col1}{RGB}{0,169,255}
\definecolor{col2}{RGB}{252,255,76}
\definecolor{col3}{RGB}{255,0,0}
\definecolor{col4}{RGB}{255,0,127}

\begin{document}
\title{Deformable MRI Sequence Registration for AI-based Prostate Cancer Diagnosis}
\titlerunning{Deformable MRI Sequence Registration for AI-based PCa Diagnosis}
%
\author{
Alessa Hering\inst{1,2} \and 
Sarah {de Boer}\inst{1} \and
Anindo Saha\inst{1} \and
Jasper J. Twilt\inst{1} \and
Mattias P. Heinrich\inst{3} \and
Derya Yakar\inst{4,5} \and
Maarten {de Rooij}\inst{1} \and
Henkjan Huisman\inst{1,6} \and
Joeran S. Bosma\inst{1,3}%
}
\authorrunning{A. Hering et al.}

%
\institute{
Department of Medical Imaging, Radboudumc, Nijmegen, The Netherlands \and
Fraunhofer MEVIS, Lübeck, Germany \and
Institut für Medizinische Informatik, Universität zu Lübeck, Germany \and
Department of Radiology, University Medical Center Groningen, The Netherlands \and
Department of Radiology, Netherlands Cancer Institute, The Netherlands \and
Department of Circulation and Medical Imaging, Norwegian University of Science and Technology, Norway’s 
}
\maketitle              
\begin{abstract}
The PI-CAI (Prostate Imaging: Cancer AI) challenge led to expert-level diagnostic algorithms for clinically significant prostate cancer detection. The algorithms receive biparametric MRI scans as input, which consist of T2-weighted and diffusion-weighted scans. These scans can be misaligned due to multiple factors in the scanning process. Image registration can alleviate this issue by predicting the deformation between the sequences. We investigate the effect of image registration on the diagnostic performance of AI-based prostate cancer diagnosis. First, the image registration algorithm, developed in MeVisLab, is analyzed using a dataset with paired lesion annotations. Second, the effect on diagnosis is evaluated by comparing case-level cancer diagnosis performance between using the original dataset, rigidly aligned diffusion-weighted scans, or deformably aligned diffusion-weighted scans. Rigid registration showed no improvement. Deformable registration demonstrated a substantial improvement in lesion overlap (+10\% median Dice score) and a positive yet non-significant improvement in diagnostic performance (+0.3\% AUROC, p=0.18). Our investigation shows that a substantial improvement in lesion alignment does not directly lead to a significant improvement in diagnostic performance. Qualitative analysis indicated that jointly developing image registration methods and diagnostic AI algorithms could enhance diagnostic accuracy and patient outcomes.

\end{abstract}
\section{Introduction}
Prostate cancer (PCa) has 1.4 million new cases each year~\cite{GCS2020}, a high incidence-to-mortality ratio and risks associated with treatment and biopsy; making non-invasive diagnosis of clinically significant prostate cancer (csPCa) crucial to reduce both overtreatment and unnecessary (confirmatory) biopsies~\cite{stavrinides2019mri}. 
MRI scans provide the best non-invasive diagnosis for prostate cancer~\cite{eldred2021population}, for which a 47\% increase in demand is expected by 2040~\cite{GCS2020}. Due to the world-wide shortage of diagnostic personnel~\cite{hricak2021medical}, workload efficiency optimization is necessary to maintain healthcare accessibility in high-income countries and improve accessibility in low and middle-income countries.

Computer-aided diagnosis (CAD) can assist radiologists to diagnose csPCa and reduce the radiology workload~\cite{winkel2021novel}, but the observed workload reduction is limited. Larger workload reduction can be achieved through autonomous operation of diagnostic algorithms. Recent advances resulted in expert-level diagnostic performance for csPCa detection algorithms using biparmetric MRI~\cite{Saha24}.

Biparametric MRI (bpMRI) consists of T2-weighted (T2W) and diffusion-weighted imaging (DWI), and the DWI is used to calculate the apparent diffusion coefficient (ADC) and typically also the high b-value (HBV) map. T2W and DWI scans are usually acquired in immediate succession in about 15-30 minutes, but slight patient movement and processes like bladder filling can lead to misalignment between sequences~\cite{kovacs2023addressing}. This misalignment results in lesion image features being misaligned between the sequences. For an accurate csPCa diagnosis, the information of both sequences are necessary to consider~\cite{weinreb2016pi}, meaning that csPCa detection algorithms have to combine information from different spatial locations when misalignment occurs. Current state-of-the-art csPCa algorithms use an early fusion strategy for the combination of the different sequences, which may lead to challenges in accurate lesion detection and characterization when the lesion image features are not well aligned~\cite{Saha24}.

To address this, misalignment in the Prostate Imaging – Cancer Artificial Intelligence (PI-CAI) dataset was manually corrected (85/1000 (8.5\%) of the test cases and 54/9107 (0.6\%) of training cases), and algorithms were subsequently trained and evaluated on these manually aligned MRI studies~\cite{Saha24}. However, manual alignment is labor-intensive, potentially undermining the efficiency gains offered by automated csPCa diagnostic methods when required during inference. Consequently, the efficacy of these algorithms in scenarios where sequences are not manually aligned remains uncertain and might be limited. 

During inference, image registration can address the issue of misaligned sequences, by providing a plausible estimation of the patient movement and deformation and thus replacing the manual alignment step. Although the prostate cancer detection research field is vibrant, there has been limited focus on the registration of prostate MRI sequences. To address the issue of global misalignment,~\cite{sanyal2020automated} proposed an affine registration approach based on prostate gland segmentation and~\cite{deVente2020deepReg} presented a rigid registration based on Mutual Information. For compensating local deformations, both~\cite{pellicer2022Reg} and~\cite{netzer2021fully}, employed the SimpleITK non-rigid B-Spline registration using Mutual Information. However, the focus of these studies was not on the evaluation of registration performance, resulting in only~\cite{netzer2021fully} examining this using the Dice Score of automatically generated prostate segmentations. In contrast, the other studies have assessed registration performance through visual examination of registered ADC images. To the best of our knowledge, only recently~\cite{kovacs2023addressing} explored the impact of image registration on prostate cancer detection performance of algorithms using bpMRI. The results show that the B-Spline registration, which is based on~\cite{netzer2021fully}, improves the overlap of manually annotated lesions as measured by the Dice score. Additionally, the performance of the downstream task of patient-level csPCa diagnosis measured by the AUROC showed a non-significant improvement from 0.76 to 0.79. These results suggest that registration is a useful preprocessing step in an automated prostate cancer diagnosis pipeline. However, due to limited sample size (only 46 positive cases in the test set) and the lack of external testing, the ability to draw definitive conclusions is hindered.

In this study, we conduct a comprehensive analysis of the impact of image registration on the clinical downstream task of case-level csPCa diagnosis, utilizing two extensive evaluation datasets. Registration accuracy is assessed through the measurement of lesion alignment across an independent dataset comprising 473 cases, each annotated with paired lesions per modality. Further, we evaluate the downstream diagnostic efficacy on an external testing set consisting of 546 cases.

\begin{figure}[t]
    \centering
    \includegraphics[width=\linewidth]{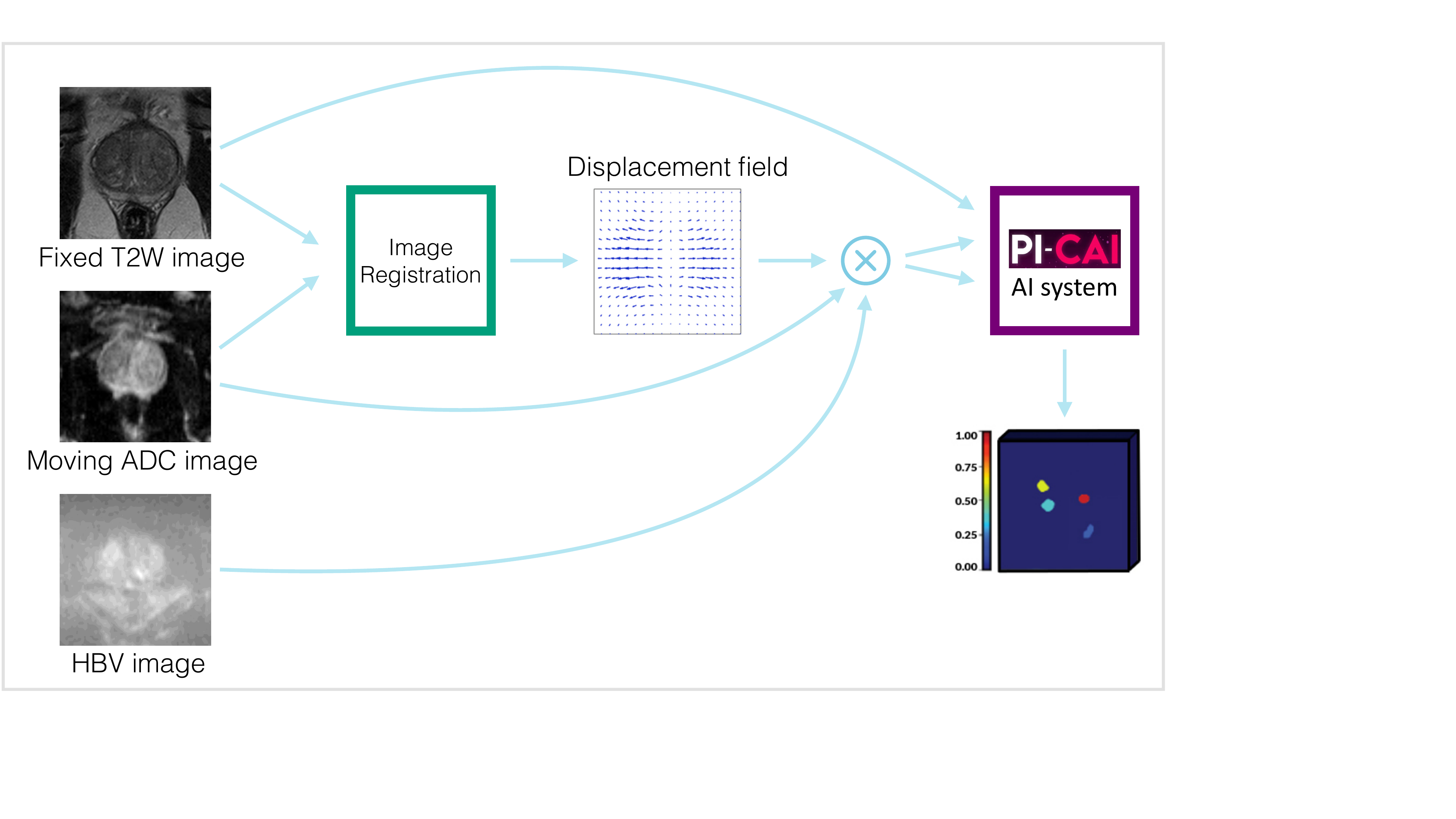}
    \caption{Overview of our method. The T2W scan is used as fixed image and the ADC map as moving image to find the displacement field using the registration method. The displacement field is applied to the ADC and HBV maps. The registered and original scans are used as input for the PI-CAI AI system (see \Cref{sec:picai_ai_system}) to detect clinically significant prostate cancer. The case-level diagnosis performance of the end-to-end pipeline is evaluated and used as a measure of effectiveness.}
    \label{fig:figure1-method}
\end{figure}

\section{Materials and methods}

\subsection{Registration}
The aim of the image registration approach is to align the DWI maps (ADC and HBV) with the T2W scan (see Figure~\ref{fig:figure1-method}). Since the csPCa detection algorithms resample the DWI maps to the T2W scan, we chose the T2W scan as the fixed image and the ADC map as the moving image for the registration. The image registration algorithm is developed in the MeVisLab framework using the RegLib. We adopt a two-step approach which consists of a rigid registration and a deformable registration. Hereby, the registration pipeline starts with robust methods with fewer degrees of freedom and moves on to more precise, but less robust methods, which require better starting points due to their higher degrees of freedom. The calculated rigid and deformable transformation are applied to both DWI maps. \\

Let $\mathcal{F},\mathcal{M}:\mathbb{R}^3\to\mathbb{R}$ denote the fixed image and moving image, respectively, and let $\Omega\subset\mathbb{R}^3$ be a domain modeling the field of view of $\mathcal{F}$. The registration method aims to compute a deformation $y:~\Omega\to~\mathbb{R}^3$ that aligns the fixed image $\mathcal{F}$ and the moving image $\mathcal{M}$ on the field of view $\Omega$ such that $\mathcal{F}(x)$ and $\mathcal{M}(y(x))$ are similar for $x\in\Omega$. \\

\paragraph{Rigid Registration} The rigid registration adopts the method of~\cite{ruhaak2017matrixAffine}. We use the normalized gradient field distance measure~\cite{HaberModersitzki2006NGF},
\begin{equation*}
   \mathcal{D}_{NGF}(\mathcal{F},\mathcal{M}(\text{y})) = \int_{\Omega} 1- \frac{\langle \nabla \mathcal{M}(\text{y}(x)), \nabla \mathcal{F}(x)\rangle^2_\epsilon}{\Vert\nabla\mathcal{M}(\text{y}(x))\Vert^2_\epsilon \Vert\nabla\mathcal{F}(x)\Vert^2_\epsilon} \, \text{d}x,
\end{equation*}
with $\langle f,g \rangle_\epsilon := \sum_{j=1}^3 (f_j g_j +\epsilon^2)$,  $\|f\|_\epsilon := \sqrt{\langle f, f \rangle_\epsilon}$, 
that focuses on the alignment of image gradients of the fixed image $\mathcal{F}$ and the deformed moving image $\mathcal{M}(y)$. The edge hyper-parameter $\epsilon > 0$ is used to suppress small image noise, without affecting image edges. The optimization problem is solved using a Gauss-Newton optimization scheme and is embedded into a multi-level scheme with two levels.

\paragraph{Deformable Registration}
We deploy the matrix-free deformable registration of~\cite{koenig2018matrixDeformable}. The deformation is defined as a minimizer of the cost function
\begin{equation*}
    \min_y \mathcal{D}(\mathcal{F},\mathcal{M}(y))+ \alpha \mathcal{R}(y),
\end{equation*}
with the normalized gradient field distance measure $\mathcal{D}_{\text{NGF}}$. To focus the registration to inside the prostate, we restrict $\Omega$ to the support of the prostate segmentation of the fixed image, which is automatically generated with the prostate segmentation algorithm provided by~\cite{Saha24}. The second-order curvature regularizer $\mathcal{R}^{curv}$ ~\cite{fischer2003curvature} enforces smooth deformation by penalizing spatial derivatives. The parameter $\alpha$ is a weighting factor of the regularizer. 
The optimization problem is solved using the limited-memory Broyden-Fletcher-Goldfarb-Shannon (L-BFGS) optimization scheme~\cite{liu1989lbfgs}. Optimization was performed in a multi-level scheme with two levels on images with successively declining levels of smoothing to guide registration from larger structures to smaller refinements. During each registration level a grid size of the displacement field of $31 \times 31 \times 31$ is used to warp the moving image using trilinear interpolation.  The deformable registration uses the output of the rigid registration as an initial starting point. Hyperparameters of the registration method were experimentally set using the first ten cases of the PI-CAI public training dataset.

\subsection{Data}
Three datasets with bpMRI scans (axial T2W, ADC and HBV (b $\geq$ 1000) imaging) for prostate cancer detection were used. For each dataset, the reference standard was set by histopathology, with clinically significant prostate cancer defined as ISUP 2-5 (intermediate to very high risk)~\cite{epstein20162014}. Informed consent was waived, given the retrospective scientific use of deidentified patient data. Scan characteristics are given in the supplementary material.

\paragraph{PI-CAI:} For csPCa detection model development, 10,207 cases of 9129 patients from 10 Dutch hospitals and 1 Norwegian hospital were used~\cite{Saha24}. Cases were acquired using 1.5 or 3-Tesla MRI scanners between 2012 and 2021 from patients with suspicion of harboring prostate cancer. Exclusion criteria included prior prostate-specific treatment, prior ISUP $\geq 2$ findings, incomplete studies, and diagnostically insufficient image quality. Manual voxel-level annotations were available for 1175 positive training cases (1323 lesions) and for an additional 892 positive training cases (1037 lesions) AI-derived voxel-level annotations were provided.

\paragraph{PCNN:} For testing of the registration algorithm, cases from the PI-CAI training set with manual voxel-level annotations per modality (T2W and ADC) were included. This selected 473 cases of 438 patients from Prostaat Centrum Noord-Nederland (PCNN) (8 hospitals).

\paragraph{PROMIS:} For external testing, 546 cases of 546 patients from 11 United Kingdom hospitals were included~\cite{ahmed2017diagnostic}. Cases were acquired using 1.5-Tesla MRI scanners between 2012 and 2015 from patients with suspicion of harboring prostate cancer. Exclusion criteria included prior prostate treatment, prior biopsies, incomplete studies, and diagnostically insufficient image quality. No manual voxel-level annotations were available.

\subsection{PI-CAI AI system}
\label{sec:picai_ai_system}
The PI-CAI AI system was developed in the PI-CAI challenge. The algorithm is the ensemble of the top 5 submissions, selected based on testing with 1000 cases. The models were trained using a dataset of 9107 cases. The algorithm uses the axial T2W, ADC and HBV scans and clinical variables (e.g. PSA density). The U-Net backbone was predominantly used, with early fusion of the scans. For additional details on the data and each of the top 5 submissions, see \cite{Saha24}. No retraining of the AI system was performed in this study.

\section{Experiments}
The aim of this study is to evaluate the effect of image registration on the clinical downstream task of case-level csPCa diagnosis. To quantify the algorithm's performance degradation under severe and extreme misalignment conditions, we artificially misaligned the T2W and DWI images in two severity steps.

On the original data, we evaluated the registration performance by measuring lesion alignment and the plausibility of the displacement field. Subsequently, we employed the csPCa detection algorithms developed in the PI-CAI challenge for the diagnostic evaluation.
For both experiments, we compare the results on three dataset variants: the original dataset, the dataset with rigidly aligned T2W and DWI scans, and the dataset with deformably aligned T2W and DWI scans.

\subsection{Impact of synthetic misalignment}
To investigate the impact of MRI sequence misalignment on the performance of a clinically significant prostate cancer detection algorithm, we conducted two synthetic misalignment tests: 

\textbf{Severe misalignment}: DWI scans were translated in the z-direction by a random selection from $\{-2,-1,0,1,2\}$ slices and in-plane by $\{-5,-4, \dots,5\}$ voxels in both the $x$ and $y$ directions.

\textbf{Extreme misalignment:} DWI scans were translated by a random selection from $-5$ or $+5$ slices in the z-direction and by $-10$ or $+10$ voxels in both the $x$ and $y$ directions.

\subsection{Registration performance}
The evaluation of registration performance was conducted using the PCNN validation dataset, chosen for its availability of lesion annotations across both T2W and ADC scans. The hyperparameters for the registration method were manually fine-tuned using only the first 10 cases from the PI-CAI Public Training and Development dataset, which did not overlap with this PCNN dataset. Therefore, the PCNN dataset serves as an independent evaluation set for assessing the registration performance. 

To quantitatively assess the quality of image registration in the absence of reference displacement fields, we employed two surrogate metrics. The Dice coefficient was utilized to quantify the overlap of lesion segmentations between T2W scans and ADC maps. Although we recognize that the Dice coefficient may not be the perfect metric for assessing the registration performance~\cite{rohlfing2011image}, its usage is justified in this context given the critical importance of accurate lesion alignment in T2W scans and ADC maps for the reliable performance of csPCa detection algorithms. The choice of the Dice score, therefore, aligns with our objective to prioritize lesion alignment in the evaluation of registration effectiveness. Smooth deformations within the prostate are important to preserve diagnostic features, therefore we evaluated the plausibility of the displacement field by examining the percentage of voxels exhibiting folding within the prostate region of the predicted deformation field. 

\paragraph{csPCa detection performance} 
Diagnostic performance is assessed using the area under the receiver operator characteristic curve (AUROC). For case-level risk estimation of significant cancer, we utilized voxel-level detection maps generated by the PI-CAI AI system on the external PROMIS test dataset. 

Additionally, we evaluated diagnostic performance using the PCNN dataset, to facilitate an evaluation of diagnostic performance in relation to the registration accuracy. We note that this dataset is not independent for the diagnostic algorithms, since this is a subset of the training data of the algorithms.

Since the PROMIS dataset contains scans with very large field-of-views with anatomical structures not present in the PI-CAI training dataset, we filtered out lesion predictions further than 3 mm away from the prostate segmentation. Following the approach used in the PI-CAI challenge, each algorithm's case-level prediction was the maximum lesion-level prediction, and the AI system's case-level prediction was the equally-weighted prediction of each algorithm.

\paragraph{Statistical analysis} \label{sec:stat_ana}
The diagnostic performance differences on the external testing set were statistically analyzed. The performance with the deformably and rigidly aligned images are compared against the performance with the original dataset. To determine the probability of one configuration outperforming another, we performed DeLong’s test~\cite{delong1988}. Multiplicity was corrected for using the Holm-Bonferroni method, with a base alpha value of 0.05. For details, see the pre-defined statistical analysis plan online \cite{hering_2024_12170878}.

\section{Results}

\subsection{Impact of misalignment}
\begin{figure}[t]
    \centering
    \setlength{\tabcolsep}{0.001\textwidth}
    \begin{tabular}{cc}
    \includegraphics[width=0.5\linewidth]{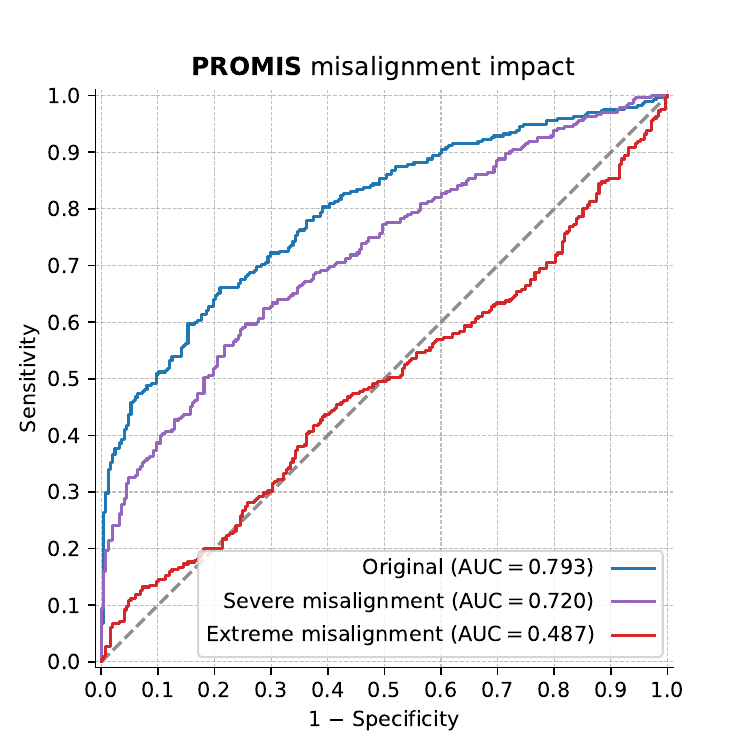} &
    \raisebox{\height}{\begin{tabular}{cc}
        \includegraphics[width=0.19\textwidth]{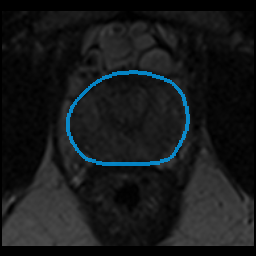} &
        \includegraphics[width=0.19\textwidth]{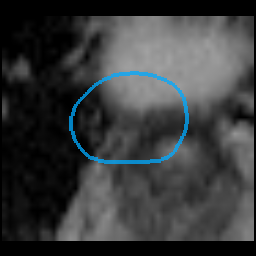} \\
        \includegraphics[width=0.19\textwidth]{figures/P-10104751_t2w_misalignment.png} &
        \includegraphics[width=0.19\textwidth]{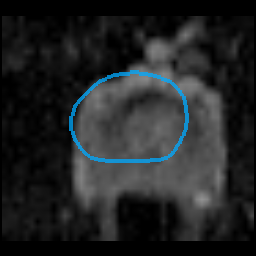} \\
        {T2W} & {ADC misaligned}
    \end{tabular}} \\
    \end{tabular}
    \caption{This figure demonstrates the impact of synthetic misalignment. (left) the diagnostic performance of the PI-CAI AI system is shown on the PROMIS dataset. When a severe misalignment is introduced, the AUC decreases from 0.793 to 0.720. In the case of extreme misalignment, the AUC further drops to 0.487.
    (right) shows the T2W and misaligned ADC (top: severe misalignment, bottom: extreme misalignment) images, with \textcolor{col1}{\rule[-0.3mm]{.3cm}{.3cm}}~prostate gland contour of the T2W scan.}
    \label{fig:misalignment-impact}
\end{figure}

In \Cref{fig:misalignment-impact}, the impact of misalignment on the performance of the clinically significant prostate cancer detection algorithm is illustrated. When a severe misalignment is introduced, the AUC decreases from 0.793 to 0.720. In the case of extreme misalignment, the AUC further drops to 0.487.

\subsection{Quantitative results}

\begin{figure}[t]
    \centering
    \includegraphics[width=\linewidth]{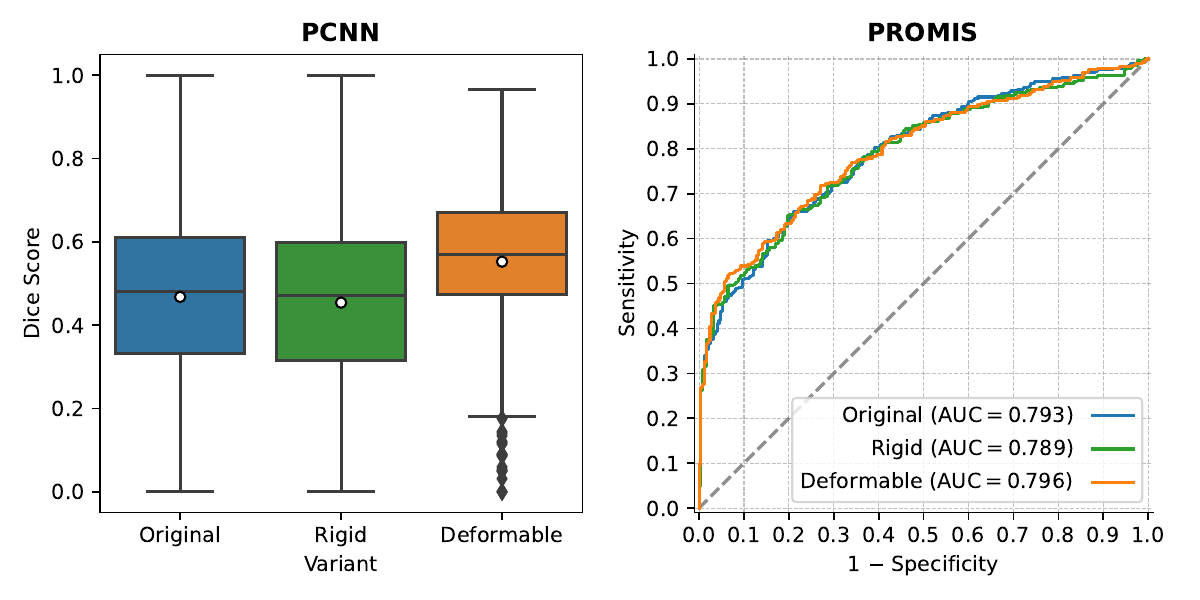}
    \caption{Quantitative registration results. \textit{(left)} Distribution of Dice scores between the lesion annotation on the T2W and ADC scans for the original, rigidly aligned, and deformably aligned PCNN datasets. \textit{(right)} Model performance for the PI-CAI AI system with the original, rigidly aligned and deformably aligned PROMIS datasets.}
    \label{fig:results}
\end{figure}

\paragraph{Registration performance}
The median Dice score improved to 0.58 with deformable registration, compared to 0.48 for the original dataset and 0.47 with rigid registration (\Cref{fig:results}).

For one case, 1\% of voxels in the prostate were folded. For all other cases, no foldings occurred in the deformation field. 


\begin{figure}[t]
    \centering
    \setlength{\tabcolsep}{0.001\textwidth}
    \begin{tabular}{cccccc}
      \includegraphics[width=0.19\textwidth]{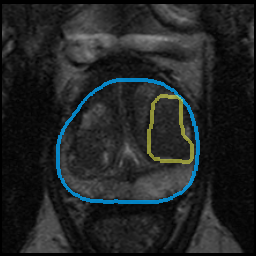}
    & \includegraphics[width=0.19\textwidth]{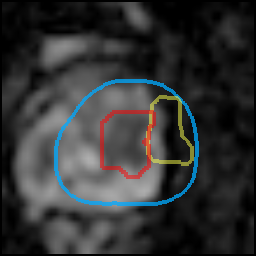}
    & \includegraphics[width=0.19\textwidth]{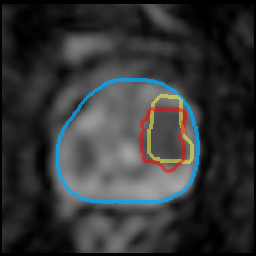}
    & \includegraphics[width=0.19\textwidth]{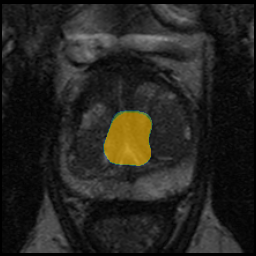}
    & \includegraphics[width=0.19\textwidth]{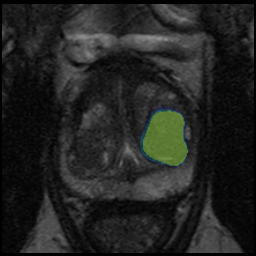}  
    & \hspace{5pt}\rotatebox{90}{\hspace{5pt}Label: ISUP 1} \\
      \includegraphics[width=0.19\textwidth]{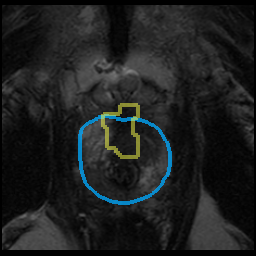}
    & \includegraphics[width=0.19\textwidth]{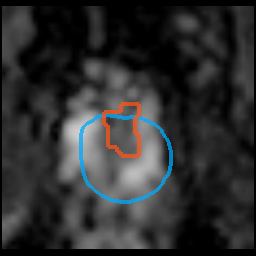}
    & \includegraphics[width=0.19\textwidth]{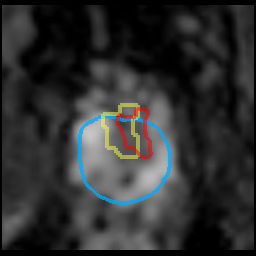}
    & \includegraphics[width=0.19\textwidth]{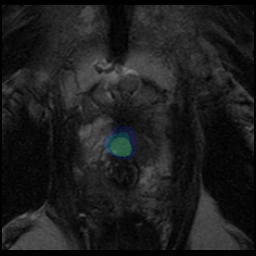}
    & \includegraphics[width=0.19\textwidth]{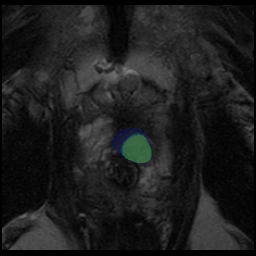} 
    & \hspace{5pt}\rotatebox{90}{\hspace{5pt}Label: ISUP 2} \\
      {T2W}
    & {ADC original}
    & {ADC deformable}
    & {PM original}
    & {PM deformable}
    \end{tabular}
    \caption{Qualitative registration results showing two exemplary cases with \textcolor{col1}{\rule[-0.3mm]{.3cm}{.3cm}}~prostate gland, \textcolor{col2}{\rule[-0.3mm]{.3cm}{.3cm}}~lesion annotated on T2, \textcolor{col3}{\rule[-0.3mm]{.3cm}{.3cm}}~lesion annotated on ADC. In the last two columns, the prediction maps (PM) generated with the original dataset and the deformably aligned dataset are overlayed on the T2W scan. } \label{fig:QualitativeResultsPCNN}
\end{figure}

\paragraph{csPCa diagnosis performance}
For the PROMIS external testing dataset, the PI-CAI AI system showed a positive yet non-significant improvement in diagnostic performance (+0.3\% AUROC, p=0.18) with deformably aligned scans compared to the original dataset. A comprehensive qualitative analysis of representative cases is given in~\Cref{app:QualitativeResults}.

\subsection{Qualitative results}
In this section, we present qualitative results of the image registration and subsequent csPCa detection algorithm. Results for the PCNN dataset are shown in \Cref{fig:QualitativeResultsPCNN}, showing the case with the largest improvement in Dice score (first row) and the largest decrease in Dice score (second row) for the deformably aligned dataset, compared to the original dataset alongside the clinical interpretation for each case. 

The first row shows the images 
with mild benign prostatic hyperplasia (BPH) in the transition zone. BPH is a benign condition, which grows over time. A typical transition zone with BPH shows so-called `organized chaos', with multiple nodules with variable imaging appearance, often with diffusion restriction and enhancement. In transition zone tumors, the typical encapsulation is lost, and the organized aspect changes to a homogeneous low T2W signal with marked diffusion restriction and vivid enhancement. In the left transition zone of this patient, an encapsulated BPH nodule is annotated in yellow on T2W, with low T2W signal intensity. On the ADC map an area with marked diffusion restriction is annotated in red. A notable discrepancy is observed in the alignment between the T2W and ADC imaging, leading to misalignment between the lesion's features on the T2W and ADC scans. 
The encapsulated nature on T2W of this nodule is a non-suspicious sign in the transition zone. 
Consequently, the PI-CAI AI system with the original scans classifies the lesion in the middle of the transition zone instead of within an encapsulated nodule more laterally due to the misalignment, which suggests a higher risk level (prediction=0.63). The deformable image registration method aligned the two modalities, and the PI-CAI AI system with the deformably aligned scans assigned a lower risk level (prediction=0.47).
Targeted biopsies revealed ISUP 1 in the left transition zone, which is an indolent prostate cancer that is often invisible on prostate MRI.


The second row shows the images for a 66-year-old man with a PSA level of 13 ng/mL and a PSA density of 0.11 ng/mL/cc. The images show a tumor suspicious area ventral in the apex of the prostate close to the anterior fibromusclar stroma (AFMS) ventral to the transition zone of the prostate. The delineation of the lesion mask was guided by the image features observed in the ADC scan and was subsequently adopted for the T2W scan as well. Upon reconsideration of the lesion segmentation with two radiologists, it appears that the extension into the AFMS is due to oversegmentation, rather than the lesion infiltrating the AFMS. As such, the model predictions capture the lesion extent very well. The prediction with the original dataset had a bit higher confidence (0.62 vs 0.56) for this positive case.
Targeted biopsy of this area revealed ISUP 2 prostate cancer.

\begin{figure}[h]
    \centering
    \setlength{\tabcolsep}{0.001\textwidth}
    \begin{tabular}{cccccc}
    \includegraphics[width=0.19\textwidth]{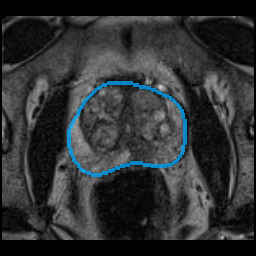}
    & \includegraphics[width=0.19\textwidth]{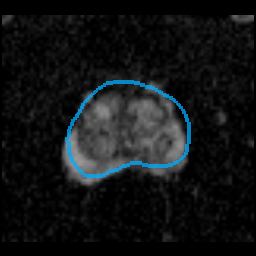}
    & \includegraphics[width=0.19\textwidth]{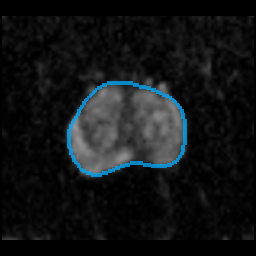}
    & \includegraphics[width=0.19\textwidth]{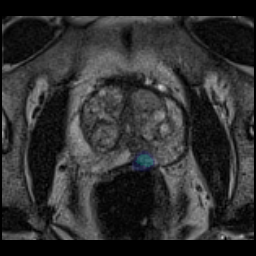}
    & \includegraphics[width=0.19\textwidth]{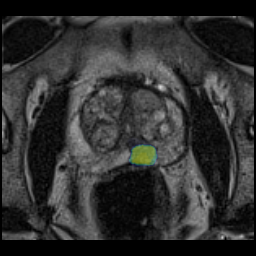} 
    & \hspace{5pt}\rotatebox{90}{\hspace{5pt}Label: ISUP 2} \\
    \includegraphics[width=0.19\textwidth]{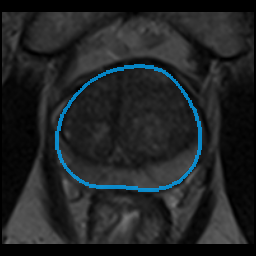}
    & \includegraphics[width=0.19\textwidth]{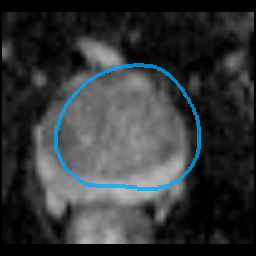}
    & \includegraphics[width=0.19\textwidth]{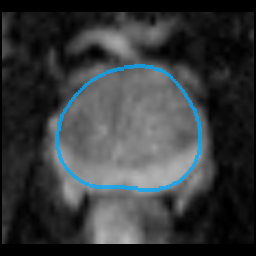}
    & \includegraphics[width=0.19\textwidth]{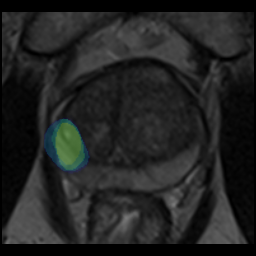}
    & \includegraphics[width=0.19\textwidth]{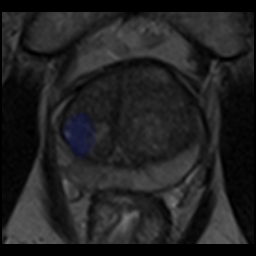} 
    & \hspace{5pt}\rotatebox{90}{\hspace{5pt}Label: ISUP 1} \\
    \includegraphics[width=0.19\textwidth]{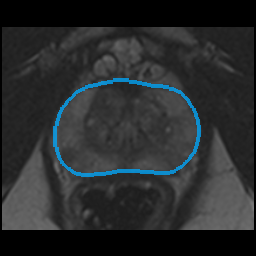}
    & \includegraphics[width=0.19\textwidth]{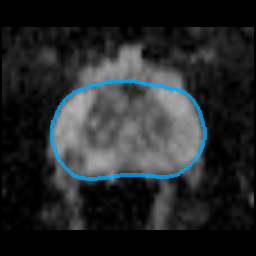}
    & \includegraphics[width=0.19\textwidth]{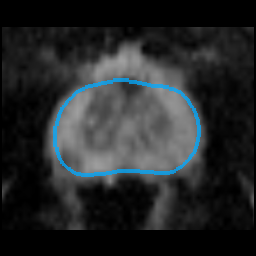}
    & \includegraphics[width=0.19\textwidth]{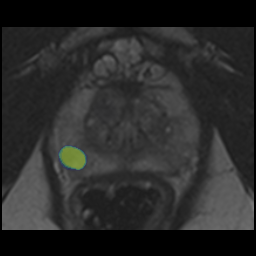}
    & \includegraphics[width=0.19\textwidth]{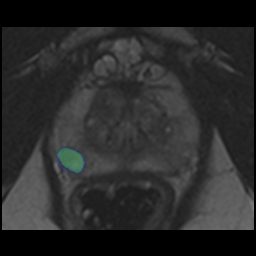} 
    & \hspace{5pt}\rotatebox{90}{\hspace{5pt}Label: ISUP 3} \\
      {T2W}
    & {ADC original}
    & {ADC deformable}
    & {PM original}
    & {PM deformable}
    \end{tabular}
    \caption{Qualitative results on the PROMIS data set. The T2, ADC, and deformably aligned ADC are shown with \textcolor{col1}{\rule[-0.3mm]{.3cm}{.3cm}}~prostate gland. In the last two columns, the prediction maps (PM) generated with the original dataset and the deformbly aligned dataset are overlayed on the T2W image. The label shows the ISUP grade, where 1 is indolent cancer (negative), and $\geq 2$ is intermediate to high-risk cancer (positive). The first two cases were selected to have the largest prediction increase and decrease for the deformably aligned dataset, compared to the original datasets, for cases with a case-level prediction above 0.3, respectively. The third case was a failure case with the deformably aligned scans. 
    } \label{fig:QualitativeResultsPROMIS}
\end{figure}

In \Cref{fig:QualitativeResultsPROMIS}, qualitative results on the PROMIS dataset are shown, which are explained in the following in more detail. 

\label{app:QualitativeResults}

The first row shows 
 images with a well-defined lesion in the left peripheral zone midprostate, with low signal intensity on T2W, and low signal intensity on the ADC map, consistent with a suspicious lesion (PI-RADS 4). The T2W and ADC map are misaligned, both in-plane and through-plane, resulting in the diffusion restriction on the original ADC map to be misaligned with the lesion features on the T2W sequence. The PI-CAI AI system identified the lesion with both variants of the dataset. With the deformably aligned dataset the algorithm confidence increased to 0.51, from a prediction of 0.36 before. Histopathological evaluation confirmed the aggressive nature of this lesion (ISUP 2).

The second row shows 
a T2W and ADC map that are misaligned in-plane. Consequently, a substantial part of the prostate on the ADC map appears outside of the prostate region of the T2W sequence. The ADC map shows diffusion restriction (low signal intensity; darker appearance) in the right transition zone midprostate. Due to the misalignment, this darker area on the original ADC map appears to be in the right peripheral zone on the T2W scan, and therefore misclassification can occur. After deformable alignment, the darker area on the ADC map aligns with the transition zone instead of peripheral zone. This is reflected in the lesion detection of the PI-CAI AI system, which predicts a lesion with confidence of 0.44 with the original scans and with a confidence of 0.18 with the deformably aligned scans. Targeted biopsies revealed ISUP 1 in the right transition zone, which is an indolent prostate cancer that is often invisible on prostate MRI. No aggressive PCa was detected.

The third row shows images 
with a small lesion in the right peripheral zone midprostate. The lesion appears as a well circumscribed area with low signal intensity (dark) on T2W images and the ADC map, suspicious for clinically significant cancer (PI-RADS 4). For this case, the deformable image registration slightly misaligned the T2W and ADC image features of the lesion, which resulted in the detection algorithm to decrease its lesion prediction from 0.49 to 0.39. Histopathological evaluation confirmed the aggressive nature of this lesion (ISUP 3).



\section{Discussion and conclusion}
In this study, we investigated the effect of image registration on the clinical downstream task of case-level csPCa diagnosis when integrated at the inference stage. Deformable registration demonstrated a substantial improvement in lesion overlap on the validation dataset (+9\% average Dice score) which is even slightly more than the one reported in~\cite{kovacs2023addressing} (+6\% average Dice score). However, since different datasets were used, a direct comparison is not possible. Moreover, the Dice score achieved with deformable registration on the validation dataset was 0.58. This performance aligns closely with the inter-rater agreement typically observed in prostate tumor segmentation. Specifically, between two radiologists independently segmenting csPCa lesions using the same modalities, the observed Dice score was 0.60 \cite{adams2022prostate158}.

Additionally, we showed a positive yet non-significant improvement in diagnostic performance on the PROMIS test dataset (+0.3\% AUROC, p=0.18) with deformably aligned scans. Our investigation shows that a substantial improvement in lesion alignment does not directly equal a significant improvement in diagnostic performance. To illustrate the impact of misalignment on the algorithmic results, we present detailed visualizations and analyses of several PCNN and PROMIS cases in~\Cref{app:QualitativeResults}. These results showed that the PI-CAI AI system demonstrated robustness to minor misalignments, particularly when these misalignments did not result in lesions being misrepresented in incorrect zones. Additionally, we anticipate a comparable number of misaligned cases in the PROMIS dataset as observed in the PI-CAI dataset, where the incidence was low. Therefore, the expected improvement in AUROC is limited. The positive yet non-significant improvement in diagnostic performance might be the result from those cases.

Our method had limitations. The deformable registration method potentially introduced unrecognized artifacts into the images which might result in worse diagnostic performance. Addressing this through retraining the csPCa algorithms to adapt to registration-induced image variations represents a promising strategy. It is crucial to note that the registration method avoided the generation of physiological unrealistic deformations. This is achieved by applying a high regularization weight to obtain smooth and plausible displacement fields.
Another critical aspect is the choice of resampling strategy. This factor considerably impacts the smoothing of ADC values, especially for small lesions, and influences the diagnostic quality of images through the effects of multiple resamples. Merging all resampling steps into one would visibly increase the quality, but is only possible in an end-to-end approach. 

The relevance of the PROMIS dataset in present-day analyses has been a subject of debate, particularly among radiologists. The diagnostic quality of MRI scans has markedly improved since the trial finished in 2015. Additionally, the PROMIS dataset contains acquired high b-value scans, while contemporary protocols calculate this based on acquired lower b-value scans, which results in less noise and better diagnostic quality. Despite these limitations, the relevance of the PROMIS dataset should not be understated. This dataset can serve as a benchmark for scenarios where access to high-end, expensive scanners is limited. This situation is a common reality for many institutions, highlighting the importance of developing algorithms that can perform well across a range of image acquisition methods.


In conclusion, our study shows that while image registration can substantially improve lesion overlap in csPCa diagnosis, it does not directly lead to a significant improvement in diagnostic performance. However, the qualitative analysis showed promising results and indicate that joint development of image registration methods and diagnostic AI algorithms could enhance diagnostic accuracy and patient outcome.



\begin{credits}
\subsubsection{\ackname} This research is funded by the European Union under HORIZON-HLTH-2022: COMFORT (101079894), Health~Holland (LSHM20103), European Union H2020: ProCAncer-I project (952159), European Union H2020: PANCAIM project (101016851), and Siemens Healthineers (CID: C00225450). Views and opinions expressed are however those of the author(s) only and do not necessarily reflect those of the European Union or European Health and Digital Executive Agency (HADEA). Neither the European Union nor the granting authority can be held responsible for them.

The PROMIS data used in the analysis for this manuscript were provided from the PROMIS study, led by University College London (UCL). PROMIS was funded by the UK Government Department of Health, National Institute of Health Research–Health Technology Assessment Programme, (Project number 09/22/67). Support was also provided by National Institute for Health Research (NIHR) UCLH/UCL Biomedical Research Centre, National Institute for Health Research (NIHR) The Royal Marsden and Institute for Cancer Research Biomedical Research Centre and National Institute for Health Research (NIHR) Imperial Biomedical Research Centre. The original PROMIS study was coordinated by the Medical Research Council Clinical Trials Unit (MRC CTU) at UCL and sponsored by UCL. The PROMIS Biobank was funded by Prostate Cancer UK (PG10-17). The PROMIS dataset and the biobank is under the research governance of the ReIMAGINE Risk Trial Management Group (funded by Medical Research Council (UKRI) and Cancer Research UK: MR/R014043/1).

\end{credits}

%
%
%
\bibliographystyle{splncs04}
\bibliography{bibliography}

\newpage

\appendix
\section*{Appendices}
\pagenumbering{roman}
\renewcommand{\theHfigure}{Sup. Mat.\thetable}
\renewcommand\thefigure{\Alph{figure}}
\renewcommand\thesection{\Alph{section}}
\renewcommand\thetable{\Alph{table}}
\setcounter{section}{0}
\setcounter{figure}{0} 

\section{Scan characteristics }

\begin{table}[h]
\caption{Scan characteristics showing the median, (95\% confidence interval) and {[}min-max{]} in voxels or mm/voxel.}
\label{tab:scan-characteristics}
\resizebox{\textwidth}{!}{%
\begin{tabular}{l|l|l|l}
                         & PI-CAI                          & PCNN                             & PROMIS                        \\
\midrule
T2W in-plane size & 640 (320, 1024) {[}256, 1078{]} & 1024 (296, 1024) {[}256, 1024{]} & 512 (256, 512) {[}256, 640{]} \\
T2W number of slices     & 21 (19, 29) {[}15, 45{]}        & 27 (20, 35) {[}15, 45{]}         & 26 (23, 38) {[}15, 94{]}      \\
T2W in-plane resolution  & 0.3 (0.3, 0.6) {[}0.2, 0.8{]}   & 0.3 (0.2, 0.7) {[}0.2, 0.8{]}    & 0.4 (0.4, 0.8) {[}0.4, 0.9{]} \\
T2W slice thickness      & 3.6 (3.0, 3.6) {[}1.3, 5.0{]}   & 3.0 (3.0, 4.8) {[}2.2, 4.8{]}    & 3.3 (3.3, 3.6) {[}0.8, 6.5{]} \\\midrule
ADC in-plane size & 128 (102, 256) {[}70, 336{]}    & 240 (114, 256) {[}108, 336{]}    & 172 (128, 172) {[}126, 256{]} \\
ADC number of slices     & 21 (19, 29) {[}11, 41{]}        & 27 (11, 33) {[}11, 41{]}         & 13 (11, 19) {[}11, 24{]}      \\
ADC in-plane resolution  & 2.0 (1.4, 2.0) {[}0.9, 2.6{]}   & 1.4 (0.9, 1.9) {[}0.9, 2.0{]}    & 1.5 (1.5, 1.7) {[}1.1, 2.0{]} \\
ADC slice thickness      & 3.6 (3.0, 3.6) {[}3.0, 5.8{]}   & 3.0 (3.0, 5.5) {[}3.0, 5.8{]}    & 5.0 (5.0, 5.5) {[}4.0, 6.0{]} \\
\bottomrule
\end{tabular}}
\end{table}

\section{Diagnostic performance}

\begin{figure}[H]
    \centering
    \includegraphics[width=\linewidth]{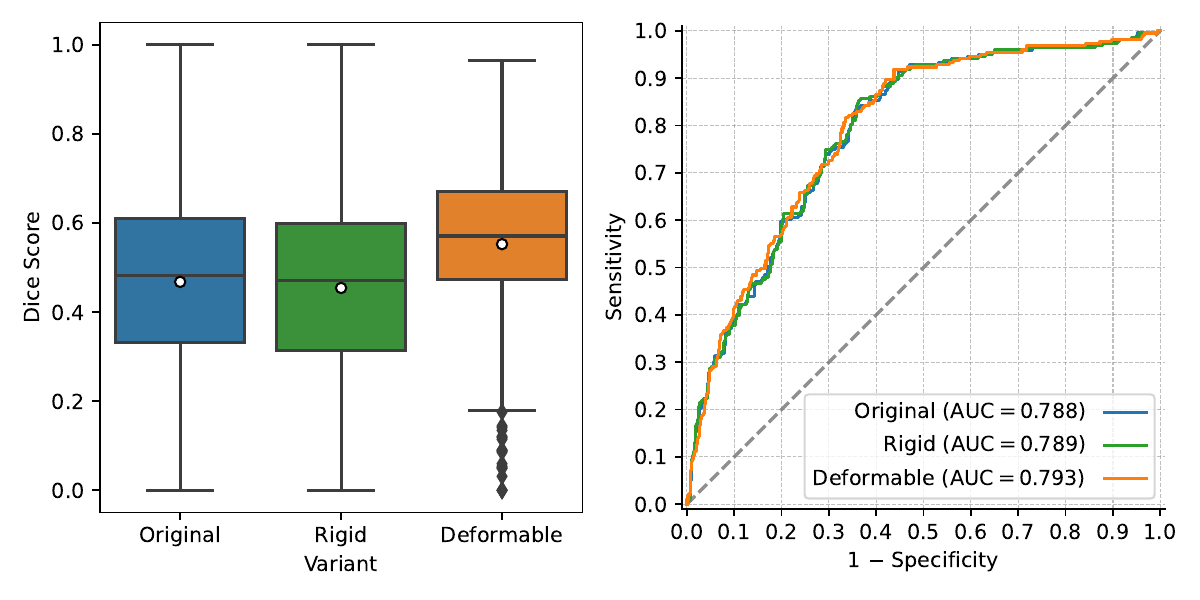}
    \caption{\textit{(left)} Distribution of Dice scores between the lesion annotation on the T2W and ADC scans for the original, rigidly and deformably aligned PCNN dataset. The median Dice score for the original dataset was 0.48, with 2.5\% and 97.5\% quantiles of the distribution of Dice scores being 0.03 and 0.90. For the rigidly aligned dataset these metrics were 0.47 [0.01, 0.82], and for the deformably aligned dataset 0.58 [0.10, 0.81]. \textit{(right)} Model performance for the Ensemble of 3 PI-CAI algorithms with the original, rigidly aligned and deformably aligned PCNN datasets. AUROC = area under the receiver operator characteristic curve.}
    \label{fig:results-app}
\end{figure}

\vspace{-10pt}
\begin{sidewaysfigure}[ht]
    \centering
    \includegraphics[width=\linewidth]{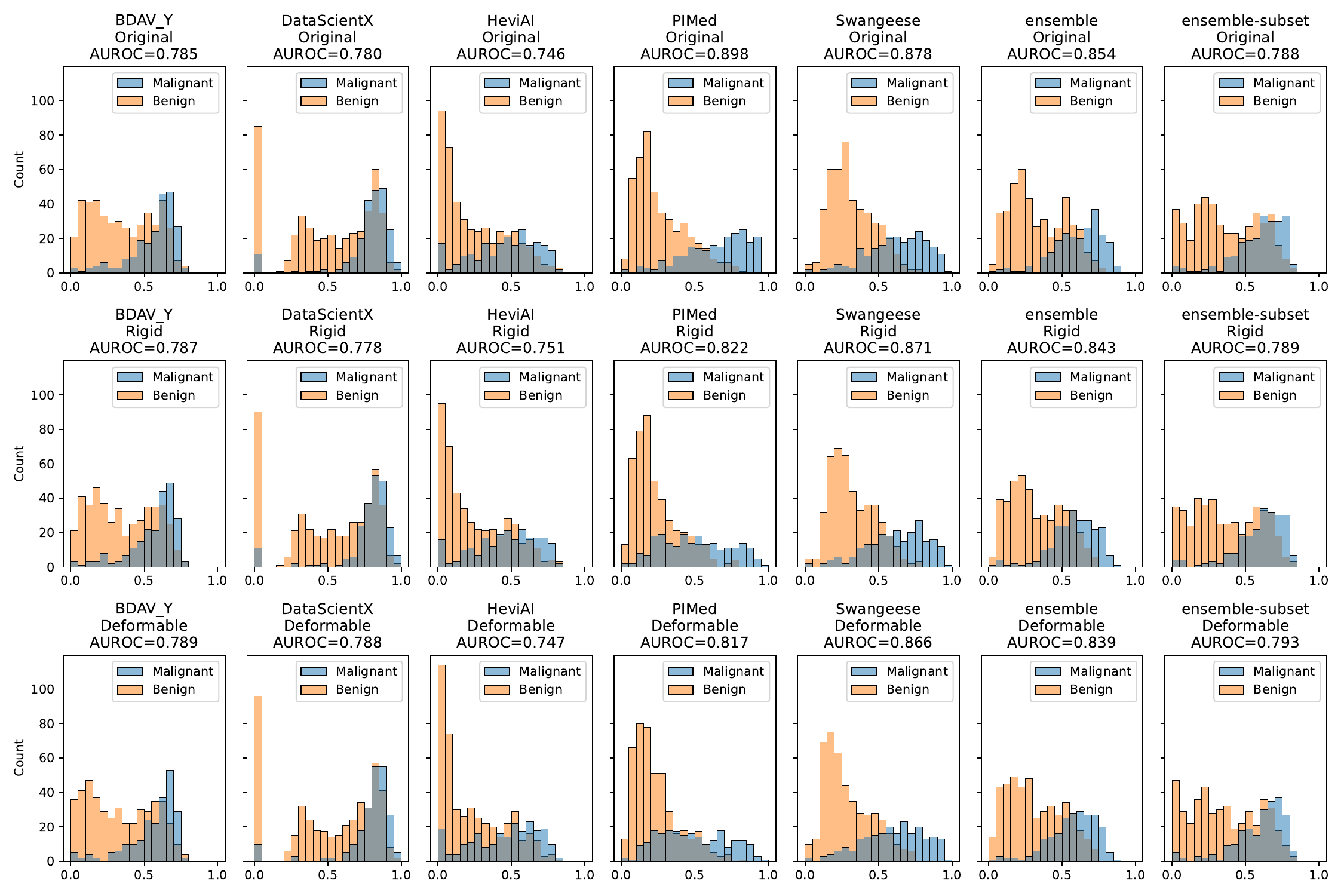}
    \caption{The prediction distribution for the PCNN dataset of each PI-CAI algorithm, their ensemble, and the ensemble of 3 PI-CAI algorithms (BDAV\_Y, DataScientX and HeviAI) ``ensemble-subset".}
    \label{fig:results-umcg-filtered-detection-maps}
\end{sidewaysfigure}

\begin{sidewaysfigure}[ht]
    \centering
    \includegraphics[width=\linewidth]{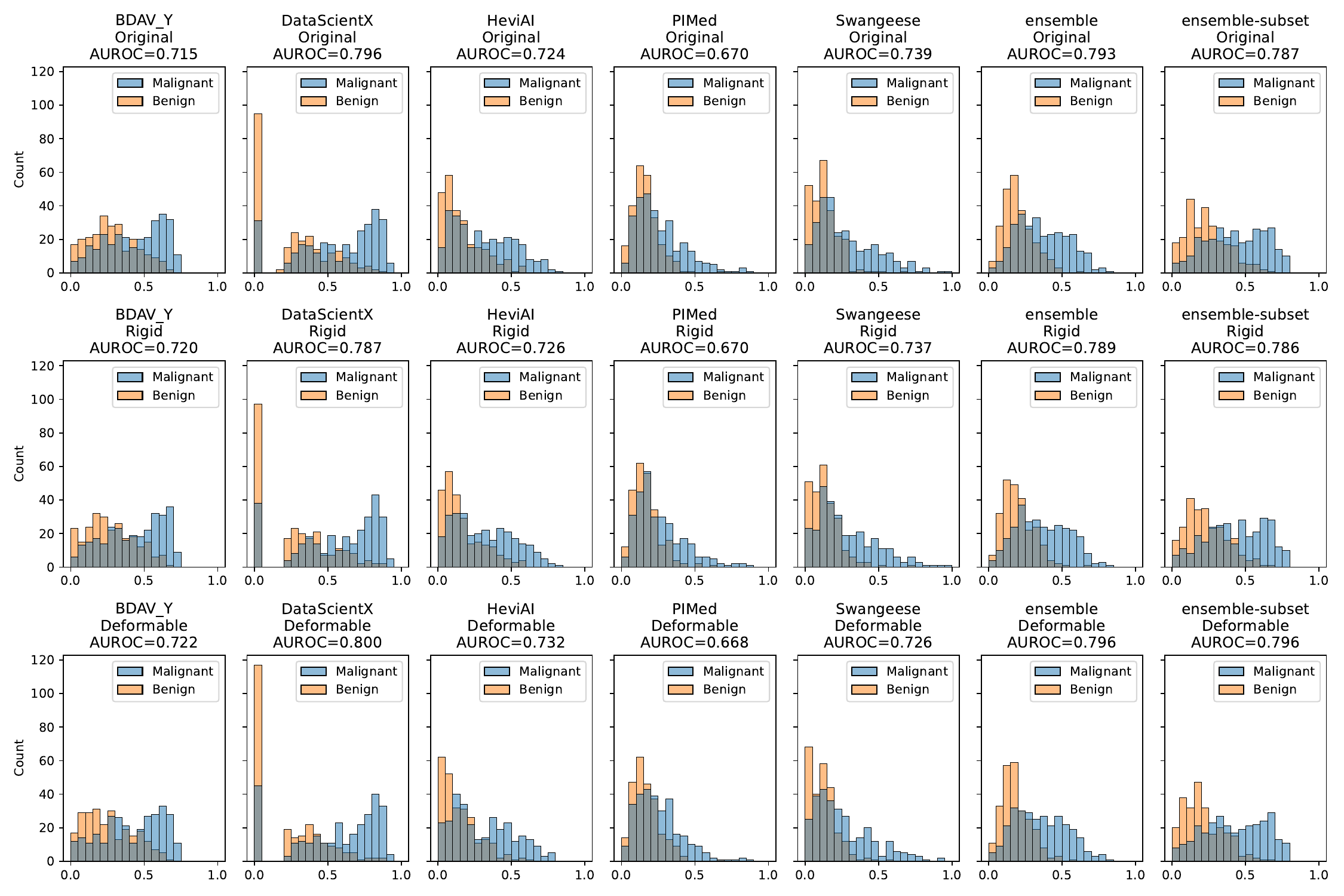}
    \caption{The prediction distribution for the PROMIS dataset of each PI-CAI algorithm, their ensemble, and the ensemble of 3 PI-CAI algorithms (BDAV\_Y, DataScientX and HeviAI) ``ensemble-subset".}
    \label{fig:results-promis-filtered-detection-maps}
\end{sidewaysfigure}

\end{document}